\documentstyle[sprocl]{article}

\input{psfig}

\begin{document}

\title{THE MORPHOLOGICAL EVOLUTION OF MERGER REMNANTS}

\author{BARBARA RYDEN \& JEREMY TINKER}

\address{The Ohio State University, Dept. of Astronomy,\\
140 W. 18th Avenue, Columbus OH 43210\\
E-mail: ryden, tinker@astronomy.ohio-state.edu}


\maketitle\abstracts{ Elliptical galaxies formed in a major
merger have a tendency to become more nearly spherical
with time, thanks to the gravitational effect of their
central black hole (or black holes). Observational results
indicate that elliptical galaxies with older stellar
populations ($t > 7.5 {\rm\,Gyr}$) have rounder central
isophotes than ellipticals with younger stellar populations.
In addition, the older ellipticals tend to have core profiles,
while the younger ellipticals have power-law profiles.
Numerical simulations of galaxy mergers indicate that
if one or both of the progenitors have a central black
hole with mass $\sim 0.2$\% of the stellar mass, then
the effect of the black hole(s) is to make the central
regions of the remnant rounder, with a characteristic
time scale of a few gigayears.
}

\section{Analytic}
In a hierarchical clustering scenario, elliptical galaxies
form by the merger of smaller stellar systems. Mergers of
equal-mass progenitors tend to form fairly flattened
systems; after violent relaxation, the ratio of the
shortest to longest axis of the merger remnant is
typically $c/a \sim 0.5$ \cite{ba92,sp00}. However,
once an elliptical merger remnant has completed violent
relaxation, its shape does not remain constant.
In the central regions of the galaxy -- well
inside the effective radius -- the morphological
evolution is driven by two-body relaxation, the
result of close gravitational encounters between
the point masses of which the galaxy is made. The
net effect of two-body relaxation is to make a galaxy
more nearly spherical with time.

Consider an idealized case of two-body relaxation in
which a mass $M$ is plunked down in an isothermal
stellar system with velocity dispersion $\sigma$.
The introduced mass will disrupt the orbits of stars
which come within a critical distance $b \sim G M / \sigma^2$.
If the mass $M$ is just another star, with mass $M \sim 1
{\rm\,M}_\odot$, then stars will have to come within
a distance $b \sim 5 {\rm\,R}_\odot ( \sigma / 200 {\rm\,km}
{\rm\,s}^{-1} )^{-2}$ before their orbits are randomized.
Thus, for a typical elliptical galaxy, stars must come
within a few stellar radii of each other for two-body
relaxation to occur, and the star/star relaxation time
is much longer than a Hubble time. If stars were the
only point masses which elliptical galaxies contained, we
would thus conclude that the effects of two-body relaxation
on the structure of ellipticals are negligibly small
so far. However, there's more to a galaxy than stars.
Most, if not all, elliptical galaxies contain central
black holes, with a mass given by the relation\cite{fm00}
$M_{\rm BH} \approx 10^8 {\rm\,M}_\odot ( \sigma / 200
{\rm\,km} {\rm\,s}^{-1} )^{4.8}$. With this black hole mass,
a star coming within a distance $b \sim 10 {\rm\,pc}
( \sigma / 200 {\rm\,km} {\rm\,s}^{-1} )^{2.8}$ will
have its orbit disrupted. Thus,
relaxation due to star/black-hole encounters will be
vastly more effective than relaxation due to star/star
encounters. The net effect of the central black hole
will be to increase the entropy of the stellar system
and to make it more nearly spherical with time.

It should also be noted that if an elliptical forms
by the merger of two progenitors, each with a central
black hole, a binary black hole may exist for many
gigayears before dynamical friction, gas dynamical
effects, and gravitational radiation will cause the
two black holes to coalesce. Three-body interactions
between a star and a bound black hole binary will
generally increase the star's kinetic energy. Thus,
binary black holes have been proposed as a mechanism
for lowering the stellar density in the central regions
of an elliptical and creating a `core' profile \cite{em91,fa97}.

\section{Observational}
Given the brevity of human life, we cannot sit
and watch for a few billion years while a post-merger
elliptical becomes rounder with time; nor can we
take the time to circumnavigate a galaxy and discover
its true three-dimensional shape at a given time.
The best we can do, to test our belief that merger
remnants become rounder with time, is to examine
a sample of elliptical galaxies and see whether
the apparent shape of a galaxy is correlated with
the time elapsed since it last underwent a major
merger.

Estimating the time that has passed since an elliptical
galaxy's last major merger is not a simple or straightforward
task. However, if the merger in question involved a
pair of reasonably gas-rich galaxies, then the merger
will be accompanied by a burst of star formation that
will leave its spectroscopic mark on the galaxy.
Terlevich and Forbes have recently compiled a catalog\cite{tf01}
of spectroscopic galaxy ages, based on a homogeneous
data set of galaxies with high-quality H$\beta$ and
[MgFe] absorption line indices. The stellar population
model of Worthey\cite{wo94} is used to assign an age
to the stellar population of each galaxy in the catalog.
For the 74 elliptical galaxies in the Terlevich \&
Forbes catalog, we searched the published literature
for isophotal fits, and found the apparent axis ratio
$q \equiv b/a$ at six reference radii: $R \equiv (ab)^{1/2} =
R_e /16$, $R_e/8$, $R_e/4$, $R_e/2$, $R_e$, and $2 R_e$.
Details of the analysis, for those who love details,
are given by Ryden, Forbes, \& Terlevich\cite{rf01}.

\begin{figure}
\psfig{figure=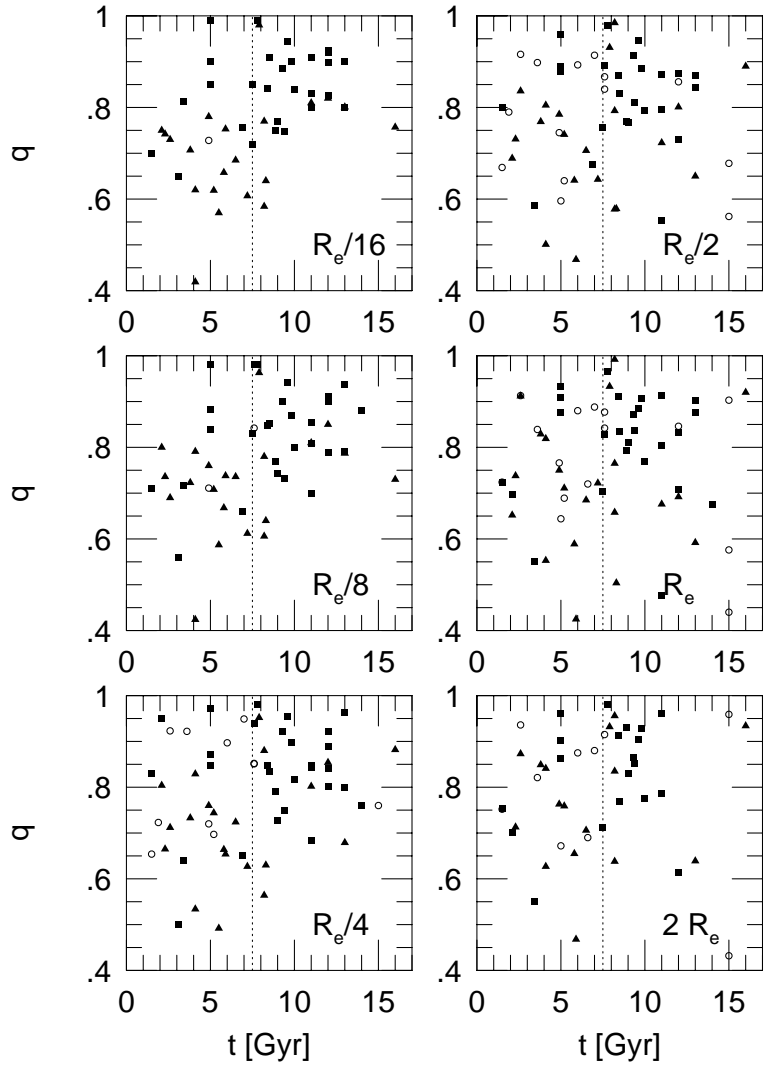,height=6.0in}
\caption{Isophotal axis ratio $q$ versus estimated age $t$
of the central stellar population. Axis ratios are measured
at $R = R_e/16$, $R_e/8$, $R_e/4$, $R_e/2$, $R_e$, and
$2 R_e$. Galaxies with core profiles are indicated by
squares, galaxies with power-law profiles are indicated
by triangles, and galaxies with unknown profile type
are indicated by open circles.}
\label{fig:forbes}
\end{figure}

Figure 1 is a plot of $q$ versus the spectroscopic age $t$
at the six reference radii. Particularly at the innermost
radius, $R = R_e/16$, there is a correlation between $q$ and
$t$, with `old' ellipticals tending to be
rounder than `young' ellipticals.
A Kolmogorov-Smirnov test, comparing the distribution of $q(R_e/6)$
for galaxies with $t \leq 7.5 {\rm\,Gyr}$ to the distribution
for galaxies with $t > 7.5 {\rm\,Gyr}$, reveals that the
distributions differ significantly, with $P_{\rm KS} = 0.00034$.

The question `Why are there so many round old ellipticals?'
is similar to the question `Why are there so many little old
ladies?' It is tempting to interpret the prevalence of little
old ladies as being due purely to the evolution of individuals, with
ladies tending to become littler as they grow older. However,
there are other effects at work as well. For instance, extremely
large ladies tend to die prematurely; in addition, today's population
of old ladies grew up when nutritional standards were lower, thus
resulting in reduced adult stature. Both these effects, neither of
which involves the morphological evolution of individual ladies,
contribute to the predominance of little old ladies over large
old ladies. Similarly, the predominance over round old ellipticals
over flattened old ellipticals is not necessarily due to the morphological
evolution of individual galaxies.

The difference in apparent shape between galaxies with young stellar
populations and those with old stellar populations is tied, in a
most intriguing manner, to the core/power-law distinction. Elliptical
galaxies with power-law profiles have luminosity densities which are
well fit be a pure power law all the way to the limit of resolution;
ellipticals with core profiles, by contrast, have densities which
show a break to a shallower inner slope\cite{fe94,fo95,la95}.
In Figure 1, core ellipticals are designated by squares, power-law
ellipticals are designated by triangles, and ellipticals of unknown
profile type are designated by empty circles. {\it Note that the `old'
ellipticals tend to have core profiles and round central isophotes,
while the `young' ellipticals tend to have power-law profiles
and flattened central isophotes.} The 29 known core galaxies in
our sample have a mean and standard deviation for their estimated
stellar ages of $t = 8.6 \pm 3.3 {\rm\,Gyr}$; the 22 known power-law
galaxies have $t = 6.9 \pm 3.5 {\rm\,Gyr}$.

\section{Numerical}

The observational results are consistent with a scenario in
which elliptical galaxies are formed in a major merger,
then evolve to become more nearly spherical with time.
However, they do not compel such a scenario -- remember
the cautionary tale of the little old ladies! Fortunately,
numerical simulations of galaxy mergers, and of the evolution
of merger remnants, can be run on timescales much shorter than
a gigayear (and, more to the point, shorter than than the lifetime of
a graduate student.)

We ran n-body simulations (with no attempt to include
gas dynamical effects) of the merger of a pair of
disk/bulge/halo galaxies. 
In our merger simulations, each progenitor has a
disk:bulge:halo mass ratio of 1:1:5.8. (The
progenitors can be thought of as a pair of S0 galaxies,
with big bulges and no gas). The disk, the bulge,
and the halo each contain 16K particles. We ran
three different merger simulations, differing only
in the mass of the central black hole assigned
to each progenitor galaxy. One simulation contained
no central black holes. In the next simulation, the
progenitor galaxies contained central black holes equal
in mass to 0.2\% of the total stellar mass (disk + bulge).
In the final simulation, the progenitors contained black
holes equal in mass to 2\% of the total stellar mass.

\begin{figure}[t]
\psfig{figure=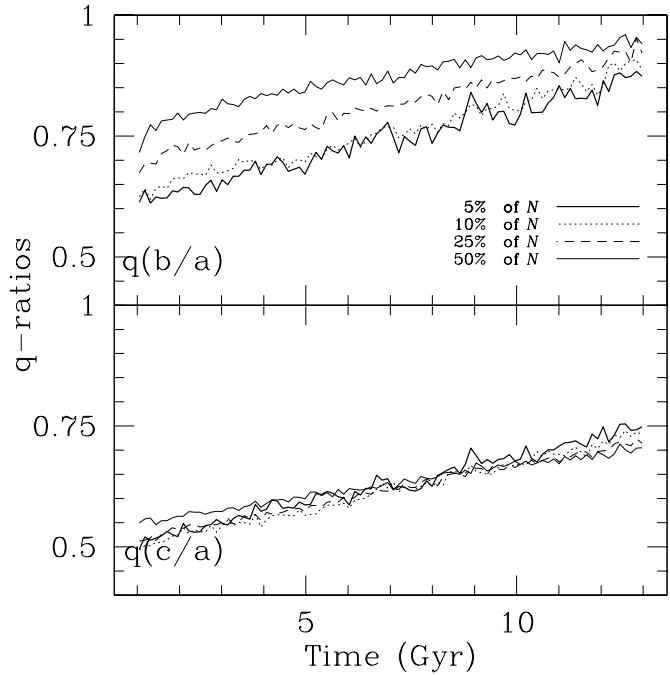,height=3.65in}
\caption{The top panel shows the evolution of the
intermediate-to-long axis ratio (b/a) of a numerical
merger remnant. THe bottom panel shows the evolution
of hte short-to-long axis ratio (c/a) for the same
remnant. In each panel, the heavy solid line indicates
the 5\% of the particles which are most tightly bound.
In this simulation, the merging progenitors did NOT
contain central black holes.}
\label{fig:nobh}
\end{figure}

We used the n-body code GADGET\cite{syw01} for all
integrations. GADGET is a tree code designed to run on distributed
memory, multi-processor computers. It employs continuously variable
timesteps which are individual to each particle. The timesteps are
computed with an accuracy parameter,
$\eta$, set to $0.02$. 
The gravitational smoothing lengths for the disk, bulge, and halo
particles were 0.08, 0.08, and 0.4 respectively. For force calculations
between any particle and a black hole, a smoothing length of 0.001 was
used. (For reference, the disk scale length of the progenitor galaxies
was 1.0.)

In the first simulation, whose results are presented in
Figure 2, the merging galaxies contained no central black
holes. Note that both the intermediate-to-long axis ratio
(illustrated in the upper panel of Figure 2) and the
short-to-long axis ratio (illustrated in the lower panel)
evolve steadily toward unity in this simulation, despite
the absence of a central black hole. This is a spurious
two-body relaxation effect, resulting from the coarseness
of our simulation; instead of being made of tens of billions
of stars, the `luminous' portions of our simulated galaxies
contain only tens of thousands of mass points.

\begin{figure}[t]
\psfig{figure=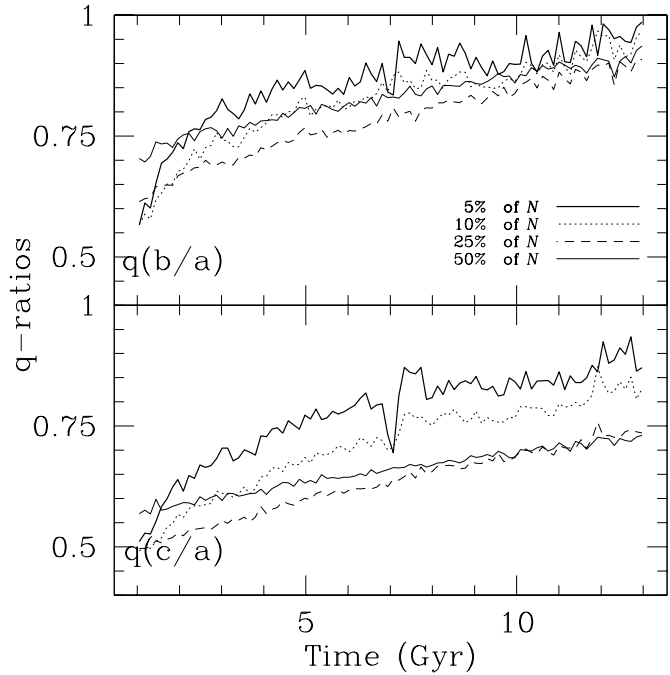,height=3.65in}
\caption{As in Figure 2, but for a simulation
in which the merging progenitors have moderately
massive central black holes (equal in mass to
0.2\% of the total stellar mass).}
\label{fig:bh002}
\end{figure}

In Figure 3, showing the evolution in shape of a merger
remnant with moderate mass black holes (0.2\% of the
total stellar mass of the progenitors), we see that the
evolution toward a spherical shape is more rapid than
in the absence of black holes. Moreover, the drive toward
a spherical shape is most rapid for the 5\% most tightly
bound particles (the heavy solid line in Figure 3) than
for the 50\% most tightly bound (the light solid line).
In short, the added black holes drive the central, most
tightly bound, regions of the merger remnant toward
a spherical shape on gigayear timescales.

\begin{figure}[t]
\psfig{figure=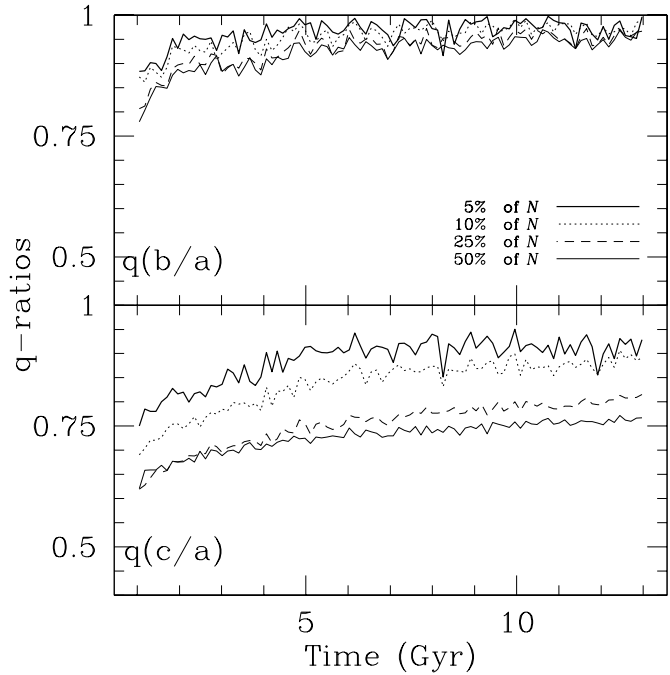,height=3.65in}
\caption{As in Figure 2, but for a simulation
in which the merging progenitors have extremely
massive central black holes (equal in mass to
2\% of the total stellar mass).}
\label{fig:bh02}
\end{figure}

Adding black holes an order of magnitude more massive,
as shown in Figure 4, dramatically shortens the time
for making the merger remnant spherical. With big black
holes, equal to 2\% of the total stellar mass, the
merger remnant rapidly becomes very nearly oblate (that
is, the ratio b/a, shown in the upper panel of Figure 4,
rapidly approaches one.) The ratio c/a, shown in the
lower panel, approaches unity more gradually. However,
we can conclude that if merging galaxies contained extremely
massive black holes, equal to one or two percent of their
total stellar mass, then merger remnants would become
nearly oblate on timescales shorter than a gigayear.
The relative scarcity of nearly circular isophotes ($q
> 0.95$) in the central regions of elliptical galaxies --
see the upper left panel of Figure 1 -- argues that
elliptical galaxies are probably not oblate in their
central regions. However, given the relatively small
size of the Terlevich-Forbes sample, this is not a
chiseled-in-stone conclusion.

More data are needed (unsurprisingly). A larger sample size
of observational data will help to pin down the relationship
among galaxy age, isophote shape, and luminosity profile type. Much
work remains to be done (also unsurprisingly). Higher-resolution
numerical simulations will reduce the ugly effects of
spurious two-body relaxation and will enable us to focus
on the physically real effects of massive black holes.


\section*{References}
\begin{thebibliography}{99}

\bibitem{ba92} Barnes, J. 1992, ApJ, 393, 484
\bibitem{em91} Ebisuzaki, T., Makino, J., \& Okumura, A. K. 1991,
Nature, 354, 212
\bibitem{fa97} Faber, S. M., et al. 1997, AJ, 114, 1771
\bibitem{fm00} Ferrarese, L., \& Merritt, D. 2000, ApJ, 539, L9
\bibitem{fe94} Ferrarese, L., van den Bosch, F. C., Ford, H. C.,
Jaffe, W., \& O'Connell, R. W. 1994, AJ, 108, 1598
\bibitem{fo95} Forbes, D. A., Franx, M., \& Illingworth, G. D. 1995,
AJ, 109, 1988
\bibitem{la95} Lauer, T. R., et al. 1995, AJ, 110, 2622
\bibitem{rf01} Ryden, B. S., Forbes, D. A., \& Terlevich, A. I.
2001, MNRAS, in press
\bibitem{sp00} Springel, V. 2000, MNRAS, 312, 859
\bibitem{syw01} Springel, V., Yoshida, N., \& White, S.D.M, 2001, New
Astron., 6, 79
\bibitem{tf01} Terlevich, A. I., \& Forbes, D. A. 2001, MNRAS, submitted
\bibitem{wo94} Worthey, G. 1994, ApJS, 95, 107

\end{thebibliography}
\end{document}